\documentclass{rmf-d}
\usepackage{nopageno,rmfbib,multicol,times,epsf,amsmath,amssymb,cite}
\usepackage[T1]{fontenc} 
\usepackage[]{caption2}
\usepackage{graphicx}
\usepackage{blindtext}
\usepackage{color}
\usepackage{hyperref}
\def\rmfcornisa{Research OR Education of Physics \hfill\rmf\ }

\def\rmfcintilla{{\it Rev.\ Mex.\ Fis.\/} }
\clearpage \rmfcaptionstyle 
\def\veck{\mbox{\boldmath$k$}}

\begin{document}

%
%
\title{Recent Developments within The Cosmic Ray Extremely Distributed Observatory (CREDO)
\vspace{-6pt}}
\author{David Alvarez-Castillo, Oleksandr Sushchov}
\address{Institute of Nuclear Physics Polish Academy of Sciences, Radzikowskiego 152, 31-342 Cracow, Poland.}
\author{Marcin Bielewicz}
\address{National Centre for Nuclear Research, Andrzeja Sołtana Str. 7, 05-400 Otwock,-{\'S}wierk, Poland.}
\author{Tadeusz Wibig}
\address{University of {\L}{\'o}d{\'z}, Faculty of Physics and Applied Informatics, 90-236 {\L}{\'o}d{\'z}, Pomorska 149/153, Poland.}
\author{Cristina Oancea, Carlos Granja}
\address{ADVACAM, 12, 17000 Prague, Czech Republic.}
\author{Noemi Zabari}
\address{Astrotectonic Ltd. (AstroTeq.ai), Juliusza Słowackiego 24, 35-069, Rzesz{\'o}w, Poland.}
\author{For the CREDO Collaboration\footnote{A complete list of authors can be found at the end of the article.}}

%
%
\author{ }
\address{ }
\author{ }
\address{ }
\author{ }
\address{ }
\author{ }
\address{ }
\author{ }
\address{ }
\maketitle
%
%
\begin{abstract}
\vspace{1em} 
%
%
This contribution presents the recent research developments within the Cosmic Ray Extremely Distributed Observatory (CREDO) in the search for resolution of various scientific puzzles, ranging from fundamental physical questions to applications like the determination of earthquake precursors. The state-of-the art theoretical, numerical and computational aspects of these phenomena are addressed, as well as recent experimental developments for detection.

\vspace{1em}
\end{abstract}
\keys{ \bf{\textit{Cosmic Ray Ensembles, earthquake precursors, machine learning, cosmic ray detectors, photon splitting, fundamental constants, quantum gravity, Lorentz invariance violation. 
}} \vspace{-8pt}}
\vspace{1cm}
\begin{multicols}{2}


\section{Introduction}

The Cosmic Ray Extremely Distributed Observatory (CREDO) represents a global effort between scientific institutes, universities, private companies, and citizens for the study of cosmic rays impacting the planet at large scales~\cite{CREDO:2020pzy}. In order to achieve this goal, CREDO aims at analyzing all the public cosmic ray data provided by all the detectors capable of registering high energy particles (including photons). This approach includes, e.g., exploring the potential of a network of smartphones working at times as radiation detectors - these devices are being used to search for extremely extended cosmic ray showers that might be as large as hundreds or thousands of kilometers. In order to improve the quality and the chances of detection, the CREDO Collaboration is also on its way to introduce more advanced detectors based on scintillators or high quality pixel cameras, as it will be described in the following sections.

Cosmic rays are high-energy particles coming to Earth from deep space. Their spectrum spans over many orders of magnitude in energy - from a few GeV to the highest energy ever observed $\sim3 \times10^{20}$ eV. The lowest-energy part of the spectrum is produced by the Sun, whereas the ultra-high energy cosmic rays (i.e. above $10^{18}$ eV) are of an extra-galactic origin, albeit unknown. Cosmic rays may be the tool to answer many questions that cover astrophysics and cosmology topics as well as more fundamental ones like the properties of space-time, the nature of dark matter or Lorentz invariance violation. Extended cosmic ray showers (or Cosmic Ray Ensembles - CRE)  that may exhibit large scale time or space correlations provide the possibility of studying those many aspects of modern physics. 

In this article, we focus on highlighting the recent developments of the CREDO Collaboration that include theoretical, numerical and experimental studies in the search for answers. This paper is organized as follows: in sections 2 to 5 we present theoretical aspects related to cosmic rays as well as the corresponding discussion of simulations. Next, a brief description of the state-of-the-art data searches in the form of machine learning is presented. Section 7 introduces the application of cosmo-seismic correlations for which the ultimate goal of the study and application of CREDO is the determination of earthquake precursors. Section 8 presents the CREDO project of cosmic ray research in Mexico where other dedicated observatories and telescopes are already established. Sections 9 and 10 present recent developments on cosmic ray detectors built from research at public universities as well as by the private sector. This article ends with an outlook and conclusions.

\section{Formation and propagation of Cosmic Ray Ensembles}
When cosmic rays interact with atmospheric gases, extensive air showers (EAS) are produced, generating cascades of secondary particles. Traditionally, the focus has been on analyzing individual EAS to estimate parameters such as energy and arrival directions of primary cosmic particles. However, CREDO pioneering paradigm shifts the attention towards Cosmic Ray Ensembles (CRE). A CRE consists of a group of at least two cosmic rays that originate from the same parent particle, irrespective of their locations in the vast expanse of the Universe. These cosmic rays within a CRE exhibit correlations in either time or space. The CREDO paradigm not only allows for the testing of existing physics scenarios, both classical and exotic, but it also possesses the flexibility to explore uncharted territories of "new physics." This innovative approach opens up avenues to delve deeper into the mysteries of cosmic rays and their fundamental properties.

Among the possible CRE scenarios, a research focus on simulation of high-energy electrons propagation, was performed. The object of the study was selected not to rule out either acceleration scenarios, corresponding to the classical models, or top-down approach, according to which the primary particles result from a lot more energetic particles, decay or annihilation. The main simulation tool, chosen for the research purposes, was a publicly available MonteCarlo code CRPropa3.1~\cite{CRPropa3_2016}, while energy loss mechanisms, responsible for CRE formation, were limited to synchrotron radiation. Postprocessing of the simulation results has been performed using a method, described in detail in~\cite{Sushchov:2022aqa}. The main goal of the investigation was analysis of the conditions favorable to observe a two-photon CRE on the Earth-sized area, aiming at estimation of the largest distances where such phenomena could come from. Our calculations demonstrate that even such a widespread electromagnetic process as synchrotron radiation in galactic magnetic fields is expected to generate CRE reaching the Earth. Moreover, in certain conditions such phenomena could be initiated at distances exceeding the Galactic scales~\cite{Sushchov2021}.

\section{Cosmic Ray Ensembles created near the Sun}

Cosmic Ray Ensembles might well be created near the Sun when an ultra-high-energy (UHE) photon interacts with the Sun's magnetic field. In order to study this possibility, the PRESHOWER code has been developed and specially adapted. The resulting showers are CRE consisting of secondary photons and particles that are extended and elongated, spanning the whole spectrum of cosmic ray energy. This effect may happen even when the primary photon is directed towards the Earth and because of the large extension of the wall-shaped shower, of the order of even hundreds of millions of kilometers with a characteristic thickness of the order of meters, the possibility of detection on Earth is relatively feasible. The relevant physical observables include  particle densities, energy spectra and geographical orientations of showers at the top of the Earth's atmosphere. Detailed analysis of this scenario can be found in ~\cite{Alvarez-Castillo:2023yzp,2022Univ....8..498P,2022JCAP...03..038D} where the characteristic \textit{galaxy-shape footprint} on of the CRE showers at the ground level derived from the simulations are presented.

\section{The search for Lorentz Invariance Violation and for the Variation of Fundamental Constants in the Universe}

CREDO can potentially discover cosmic ray showers where effects from Lorentz Invariance Violation (LIV) can be revealed. This effect is motivated by quantum gravity theories that predict a different dispersion relation for photons, for instance in~\cite{Jacobson:2005bg} the authors predict
\begin{equation}
E_{\gamma}(\veck)=\sqrt{\frac{(1-\kappa)}{(1+\kappa)}}\left|\veck\right|,
\end{equation}
where $\veck$ is the photon momentum and $\kappa$ characterizes the deviation. Alternatively, some theoretical extensions of the Standard Model of Particle physics introduce LIV and CPT violations through spontaneous symmetry breaking caused by background fields which produce some preferred frame effects, resulting in modified dispersion relations  for matter and light as well~\cite{Colladay:1998fq}. 

Interestingly, some other fundamental constants may have changed their value in the early Universe~\cite{AlvarezCastillo:2022wtk}. Some string theories predict that the gravitational constant $G$ may have a dependence on the shape of a potential for the size of internal space scale-lengths in the extra dimensions introduced by definition~\cite{Ivashchuk:1988bs}. The so-called \textit{Dirac large number hypothesis} introduces the notion that the variation of $G$ allows for an explanation of the weakness of the gravitational force with respect to the others~\cite{Dirac:1938mt}. All in all, the fundamental constants might be related to each other, as considered in the definition of the \textit{Planck Energy Scale} and its associated \textit{Planck mass} $G=\hbar c /M^{2}_{P}$, and \textit{Planck length}  $l_{p} = \sqrt{\hbar G /c^{3}}$, thus the dependence on their possible variations.

\section{Photon splitting}

It is expected that Cosmic Ray Ensembles can be created in the vicinity of compact objects like Magnetars and Neutron Stars,
objects bearing very strong magnetic fields~\cite{1979JPhA...12.2187S,1970PhRvL..25.1061A,1997ApJ...476..246H}. The production mechanism consists of photon splitting $\gamma \rightarrow \gamma
\gamma$, a quantum electrodynamic process without threshold, which becomes relevant once the magnetic field falls above the critical value $B_{crit}=4.41\times 10^{13}$ G up to the onset of magnetic one-photon pair production, $\gamma \rightarrow e^{+} e^{-}$. Importantly, the photon splitting rate can become large enough during a photon's propagation through the neutron star magnetosphere before 
the pair production threshold is crossed, suppressing the production of secondary electrons and positrons in pair cascades. Under this condition, 
we are able to compute the photon splitting process, which takes into account the dipole field and curved space-time geometry of the neutron star magnetosphere. When the splitting rate becomes large enough, splitting can take place during a photon's propagation through the neutron star magnetosphere before the pair production threshold is crossed. Of great interest is to assess the probability of this phenomenon to occur for very energetic particles, like those of the order of $10^{20}$ eV, and whether or not very strong magnetic fields are required. These kinds of both theoretical and numerical studies are at the moment being carried out within the CREDO collaboration.

\section{Machine Learning approaches to searches within CREDO data}

\begin{figure}[H]
\centering
\includegraphics[width=0.2\textwidth]{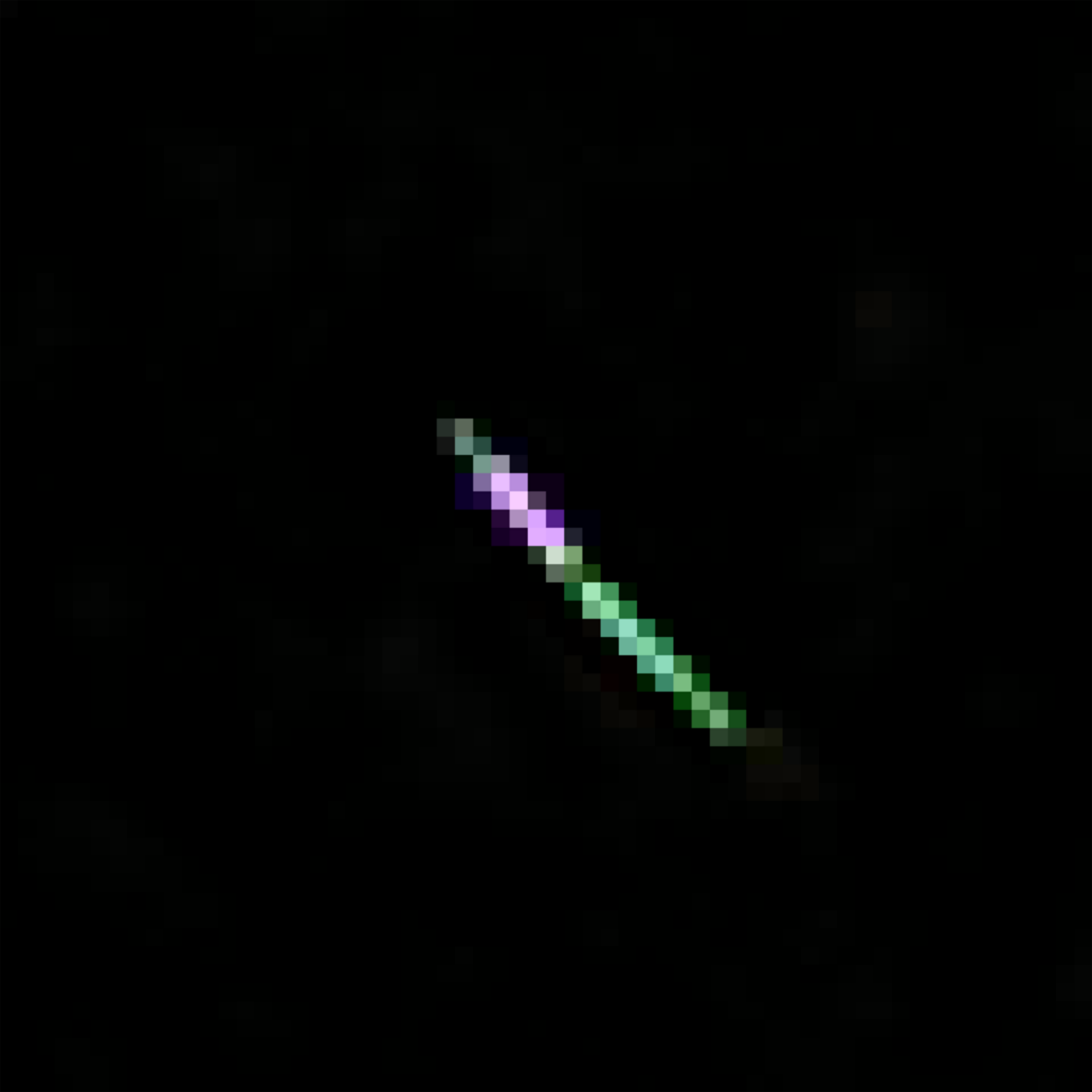}
\includegraphics[width=0.2\textwidth]{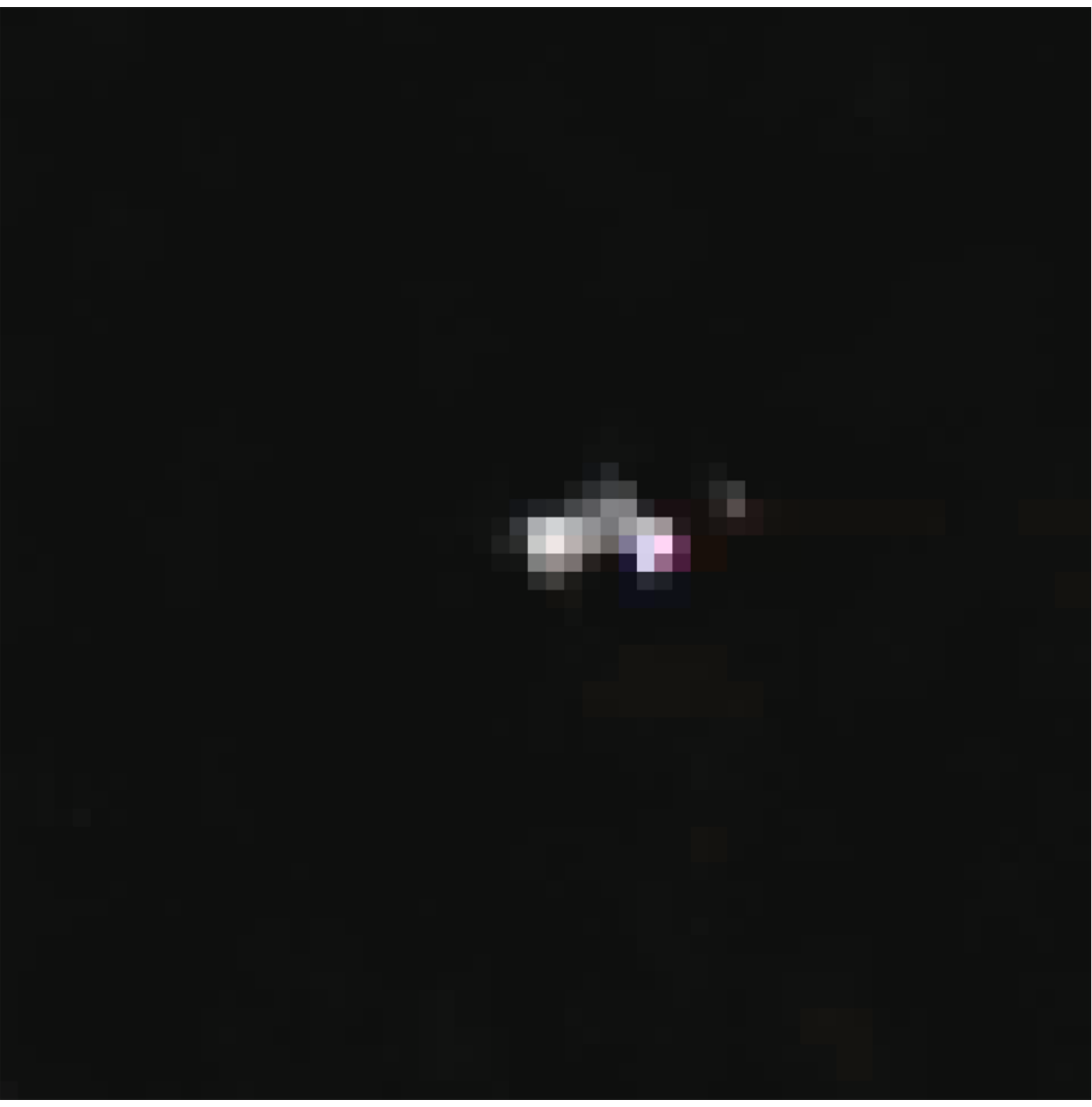}
 \caption{Particle detection by a smartphone. The images are transferred to a supercomputer in order to be analyzed for particle identification, usually by Machine Learning techniques.}
 \label{fig:detection}
\end{figure}

One very important aspect of the CREDO Collaboration is the identification of detected particles. Figure~\ref{fig:detection} shows typical images of a smartphone detection of a particle in the form of a trace. All the images are collected in a supercomputer in order to be analyzed. Typical methods include Machine Learning techniques. One of the main challenges of this approach is the need to avoid enormous training data sets that would result computationally expensive. That is precisely the case of Monte Carlo (MC) generators like CORSIKA~\cite{Heck:1998vt}. A recent study has shown that the Nakamura-Kamata-Greisen (NKG) distribution can capture the main features of full-fledged simulations. The result of introducing this method is that a typical sample of $10^{6}$ muons may take about a minute in contrast to a computationally expensive MC simulation of the order of a few days~\cite{10019257}.
\section{Cosmo-seismic correlations}

One very practical and novel possibility of exploitation of the CREDO net of detectors is the search for earthquake precursors. The main hypothesis is that a disturbance of the Earth's magnetic field before earthquake occurrence can potentially disturb or alter the flux of charged particles contained in cosmic ray showers. The large distribution of smartphones and any other dedicated cosmic ray detector allows for continuous monitoring, with the advantage that any limitation in the time resolution of time correlated particles would not play an important role. The ultimate goal is the 
development of an early warning system against earthquakes. In order to achieve this, the main strategy should include various types of data, from seismological measurements to atmosphere ionization levels to satellite observations of the Earth. The recent observation of correlations between cosmic rays and earthquakes which appear with a periodicity similar to the solar cycle opens a new perspective for identifying new earthquake precursors observable at large scale, over the entire planet~\cite{Homola:2022mry}. The data considered there includes the public Pierre Auger observatory scalar data set as well as a few stations of the Neutron Monitor Database (NMDB), and the U.S. Geological Survey (USGS) database earthquake data. Among other effects that may be incorporated into these studies are the tidal forces exerted by the Sun and Moon over the Earth. Alternatively, different methodologies, like a Bayesian analysis that searches for changes in modulations of the various aforementioned quantities may serve as an earthquake precursor indicator. All in all, geophysical modeling of the Earth dynamics and its magnetosphere is utterly relevant for this study.
 
\section{Astroseismic project - CREDO MEXICO}

As part of the CREDO program, analyses that search for correlations between the level of cosmic radiation and the average level of earthquakes are being carried out. These analyses so far show that such correlations occur at a global scale. That is, a comparison of long-term cosmic ray measurements from a specific observatory and global averaged earthquake data showed a certain degree of correlation~\cite{Homola:2022mry}. Further analyses of this kind that look for correlations of other physical quantities are underway, still based on global earthquake data and information from a selected cosmic ray observatory but incorporating new data sets.

The main question is whether such correlations exist (and how feasible is to find them) when we consider local earthquake data from a specific small area (e.g. 10,000 km$^2$) and cosmic ray data from the same area. For this purpose, we should choose a seismically active region, which is also densely populated and has academic or research facilities in its area. We propose as such a region of central Mexico that include the cities of Mexico and Puebla. This region is very seismically active, including major earthquakes. The two mentioned cities are about 100 km apart and are separated by an active volcano. The entire greater region is surrounded by mountains, many of which are extinct volcanoes. Therefore, an extensive network of seismological measuring stations has been actively operating in the entire region for several decades, based on the National Seismological Service, an organization that studies and records earthquake activity.

As part of the new CREDO-Mexico project, we intend to place several medium-sized cosmic ray detectors in this area, which would be located in academic units (e.g., University of Puebla and UNAM in Mexico City). Such a location would ensure the possibility of long-term data collection (at least several years) from several independent measurement points scattered in this area. The data obtained in this way can be compared with those collected simultaneously from many data points concerning the seismology of this area. The main goal of this program would be to try to find local correlations between these two phenomena. In addition, an analysis of historical data is also envisaged. In this area of Mexico, The High Altitude Water Cherenkov Observatory (HAWC)~\cite{Springer:2016xzh}, a gamma-ray and cosmic ray observatory located on the flanks of the Sierra Negra volcano, will provide complementary data of interest to us. Near this location, The Large Millimeter Telescope~\cite{2020SPIE11445E..22H} can be found as well. Building a network of such observation points is necessary so that the collected data are not dependent on very local fluctuations, and they will represent mutually referenced data. A second location for this measurement program is being considered in Chile, where we also have a group of collaborators, with the advantage that this area would be relatively close to The Pierre Auger Observatory \cite{PierreAuger:2015eyc}.

\begin{figure}[H]
\centering
 \includegraphics[width=0.5\textwidth]{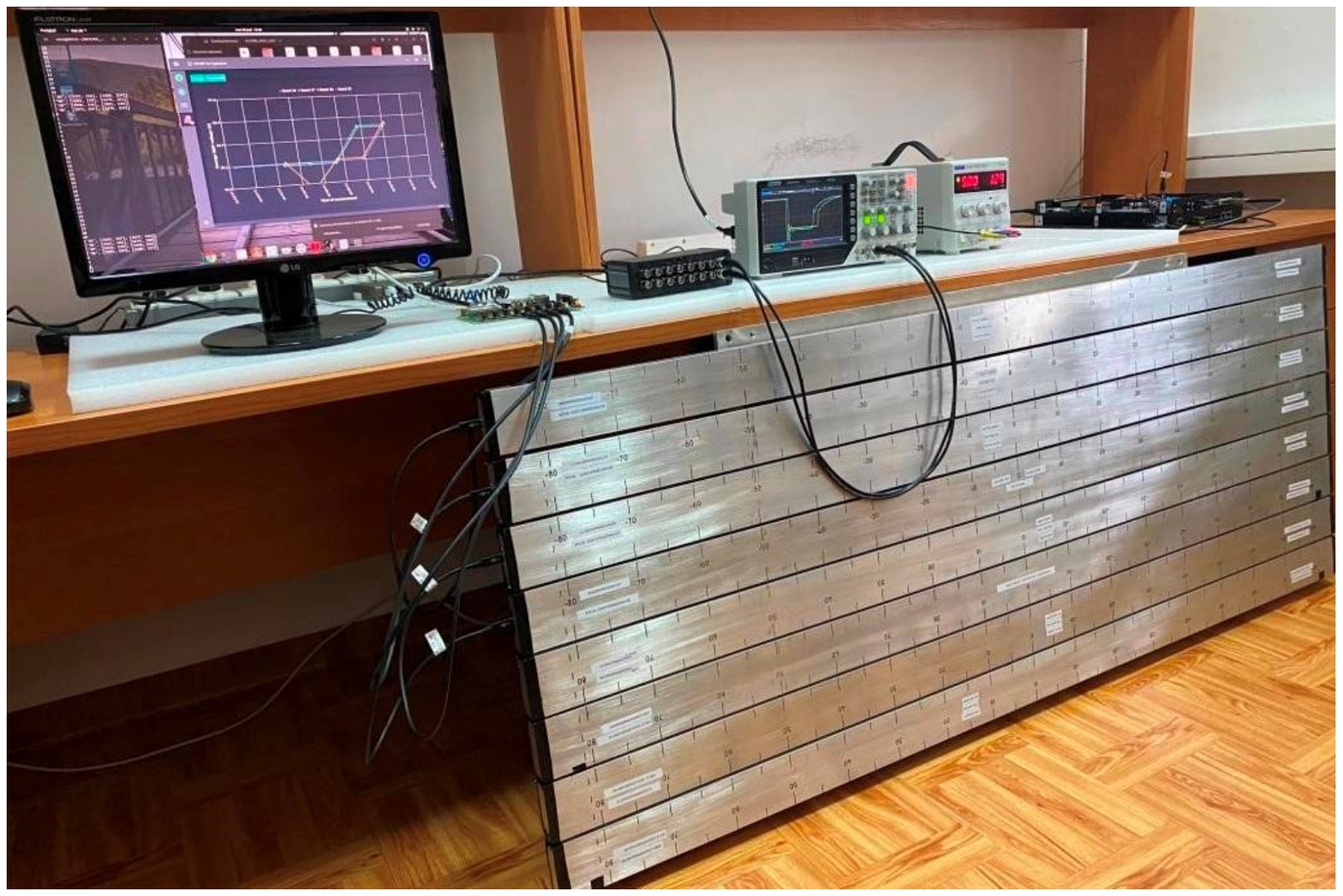}
 \caption{Real photo of the MCORD section during test measurements.}
 \label{fig:imgMCORD2}
\end{figure}

For the construction of cosmic ray detectors (detectors of muons that come from atmospheric cascades generated by primary cosmic ray particles in our atmosphere), it is planned to use the existing MCORD (Mobile Cosmic Ray Detector)~\cite{Bielewicz:2021olh,electronics12061492,2020EPJWC.23907004B} detector, which was developed at the National Center for Nuclear Research (NCBJ), Poland.
\begin{figure}[H]
\centering
 \includegraphics[width=0.5\textwidth]{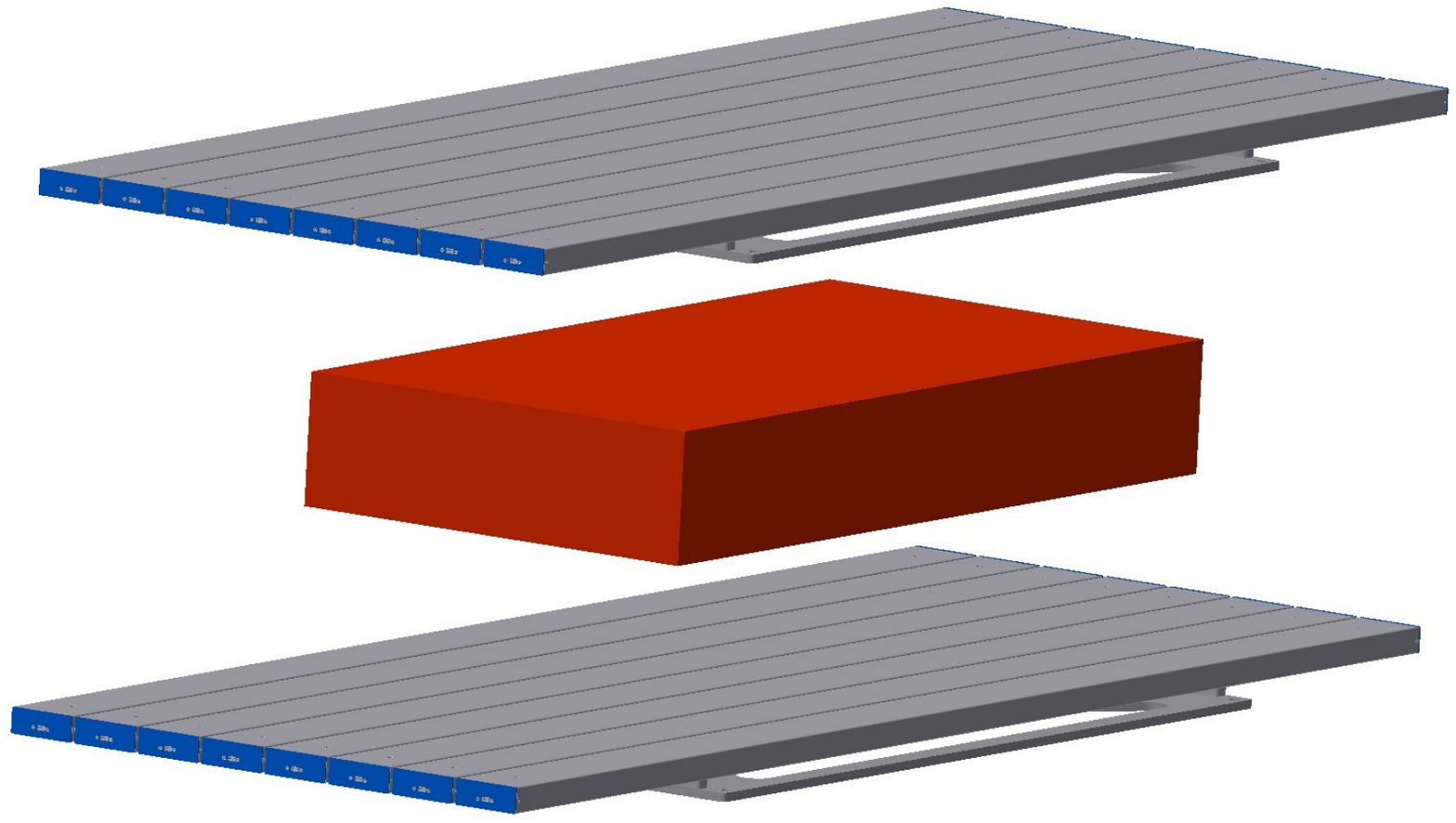}
 \caption{Two MCORD sections, each containing 8 scintillators (grey) placed one above the other. Between them we can place (red) the analyzed object, particle absorber or leave this space empty.}
 \label{fig:imgMCORD1}
\end{figure}
%
The detector units would have an effective area of about 1m$^2$, and would consist of one or two independent layers, placed one above the other (see figures \ref{fig:imgMCORD1}- \ref{fig:imgMCORD2}). Each layer (called MCORD section) consists of eight independent longitudinal scintillators (length over 1.5 m, width approx. 7 cm, thickness approx. 2.5 cm) with a built-in single optical fiber and independent reading of light from both ends. The Readout system would be based on silicon photomultipliers (SiPM), each equipped with its own controllable power supply, temperature sensor, and amplifier. Each received signal would be checked for the minimum value of the amplitude and the correct (time window) coincidence of the signal from both ends of the given scintillator. In the case of using two layers (two MCORD sections), the coincidence of signals from both layers would also be analyzed. The second variant would also give the possibility of identifying the direction of the incoming signal (the direction of the muon's flight). Each section will have its own independent reading and supervision system (Detector Control System). The physical signal would be directed to an analog-to-digital converter and then analyzed by a system based on FPGA type electronics. The system designed in this way will allow us to easily install it at a given measurement point. Data will be collected reliably and with very high efficiency (99\%).
A group of people from several scientific institutions from Poland, Mexico and Chile (under the CREDO program) are interested in this project and are currently applying for funding.

\section{The CREDO-Maze Project}
The original concept of the CREDO Project was to use mobile phones, specifically their camera CCD matrix, as detectors of cosmic ray particles. They could be used to record Extensive Air Showers and, from their global distribution, to search for Cosmic Ray Ensembles.

This idea was developed by a group from the University of {\L }\'od\'z into the idea of a network of local small-scale big-band apparatus organized for the same purpose. The project has been named CREDO-Maze, referring to the co-discoverer of Extensive Air Showers Roland Maze, who built the first cosmic ray array on the roof of the \'Ecole Normale Sup\'erieure in Paris in the late 1930s. 
An essential element of the Project is to use the network of high schools as a logistical base and to involve young people in operating and monitoring the performance of their own array as an element of the network structure. A side effect, perhaps even more important than the physical task itself, is to give young people a unique opportunity, unavailable in any other way, to experience modern methods of doing science and to participate in science as it is in the real world.

Building a small shower array today is not a big technical problem.  However, the designers of the CREDO-Maze had to minimize the cost while meeting the requirements of the physical problem, with the aim of making the machine available to as many schools as possible.

\begin{figure}[H]
    \centering
 \includegraphics[width=0.4\textwidth]{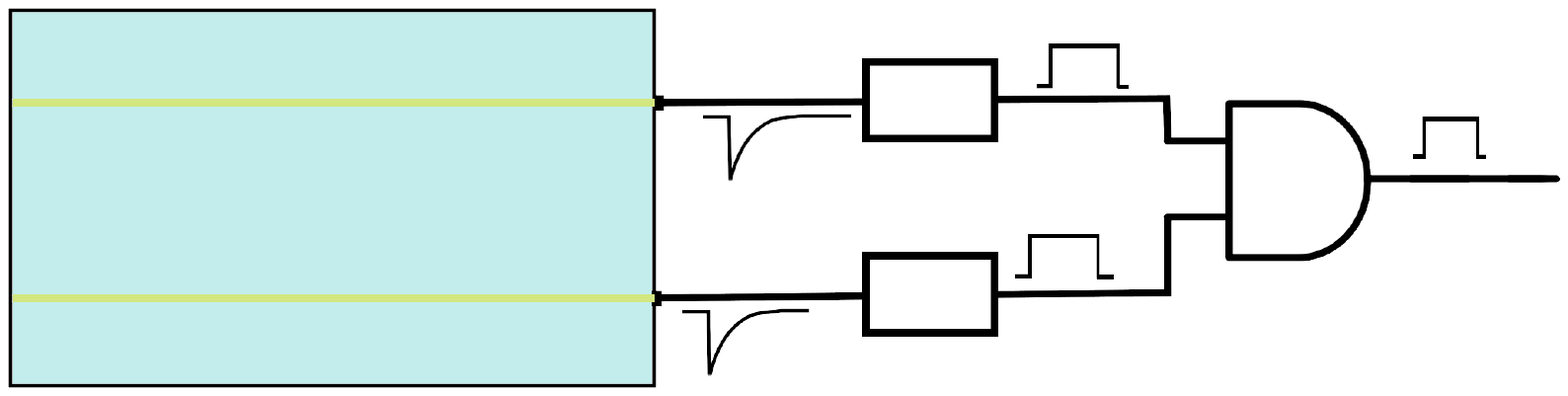}
    \caption{Logic of single CREDO-Maze detector}
    \label{fig:my_label1}
\end{figure}

\begin{figure}[H]
    \centering
   \includegraphics[width=0.5\textwidth]{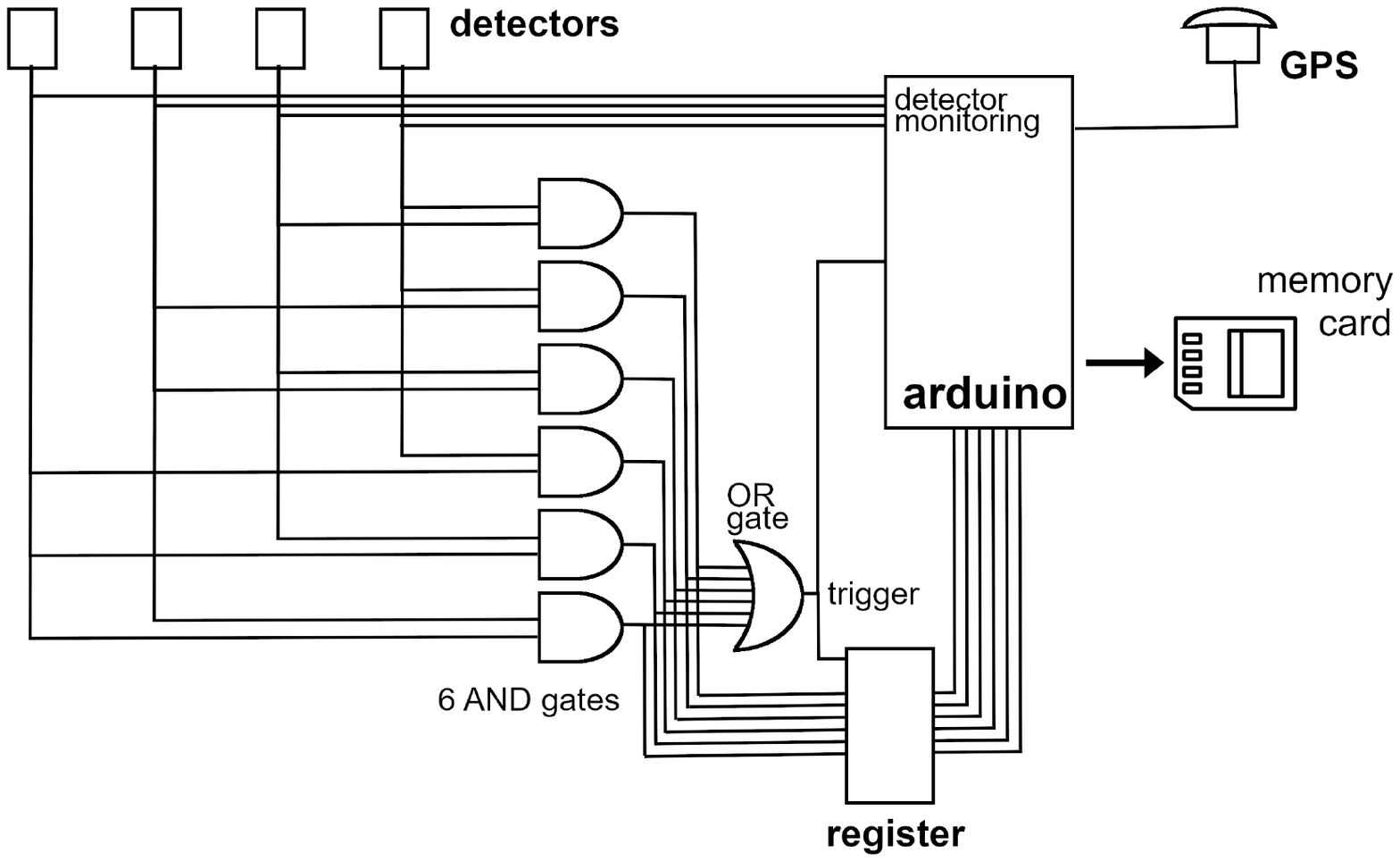}
    \caption{Scheme of a local detection station}
    \label{fig:my_label2}
\end{figure}

Thus, a school apparatus consists of four 0.02 m$^2$ scintillation detectors equipped with two optical fibers, simultaneously observed by a silicone photomultiplier working in coincidence. They are connected to a central unit whose task is to identify shower events and record information about them on typical memory cards, together with the time of the event measured by a GPS chip (with an accuracy of 1 ms). The station is controlled by a simple Arduino module and powered by a small power supply or even a power bank for a while. The apparatus does not require a dedicated, continuously running computer, so the cost of a complete set supplied to the school is comparable to that of a standard desktop computer. The data from the card is read and analyzed off-line and, after verification, transferred to the project database and, in the future, to the main CREDO project database (see figures \ref{fig:my_label1}- \ref{fig:my_label2}).
%
%
Simulations show that a single station will record showers initiated by primary cosmic ray protons with energies above 10$^{13}$eV, and such events are expected at a rate of several per hour, depending on the type of coincidence. At the time of writing (May 2023), we have completed the first pilot phase of work, during which we have developed the final version of the detectors and central station and tested their usability and stability. We have already installed the devices in four high schools in the city, and work is underway to calibrate them, connect them to the Project network and, in the near future, transfer data to the CREDO Project databases.
\section{CREDO Interplay with the private sector}
The CREDO Collaboration is open to citizen science as well as to collaboration with the private sector, with some of the workers providing scientific support. Below are a couple of cases.
%
\subsection{ADVACAM}
\begin{figure}[H]
\begin{center}
 \includegraphics[width=\linewidth]{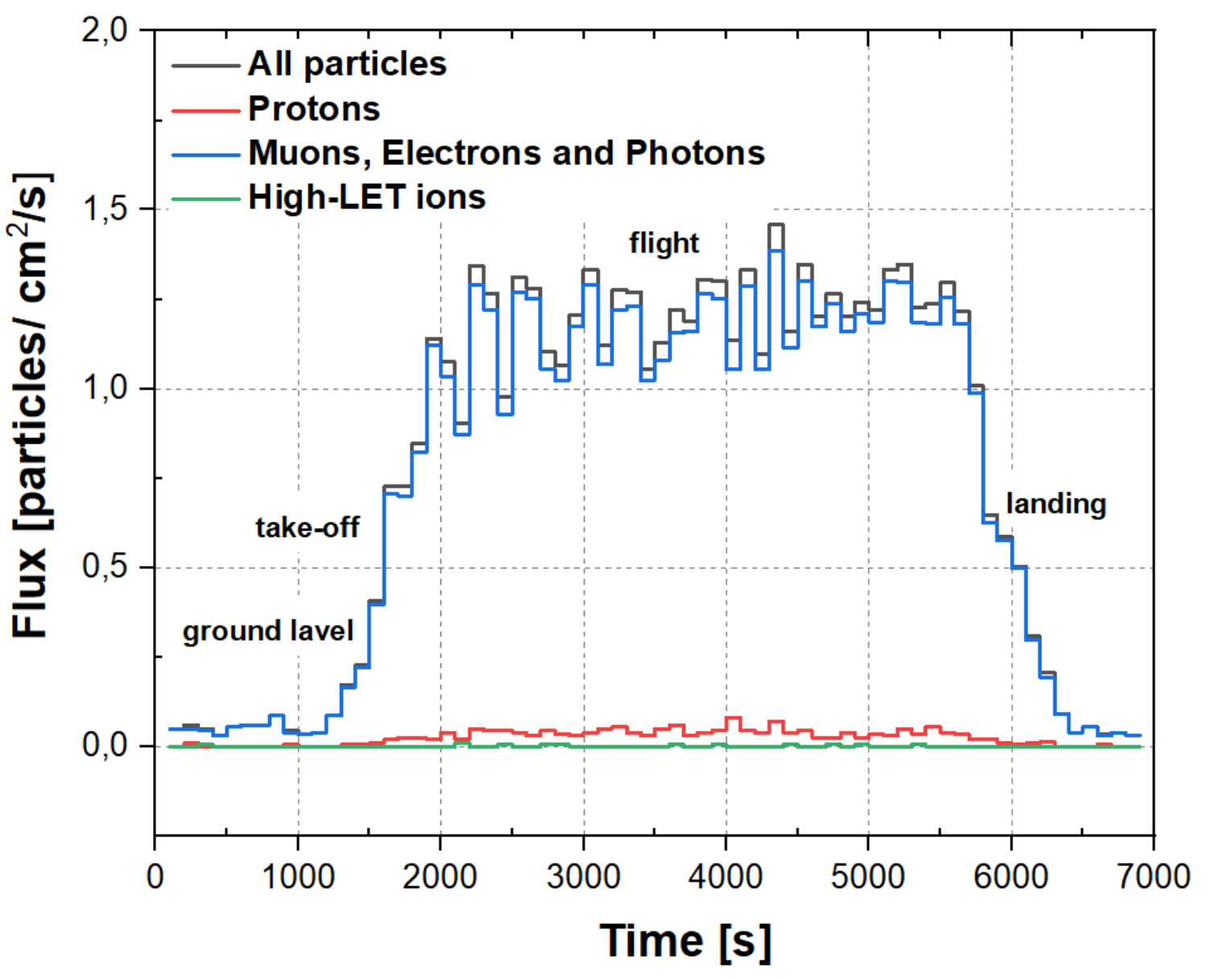}\\
 \includegraphics[width=\linewidth]{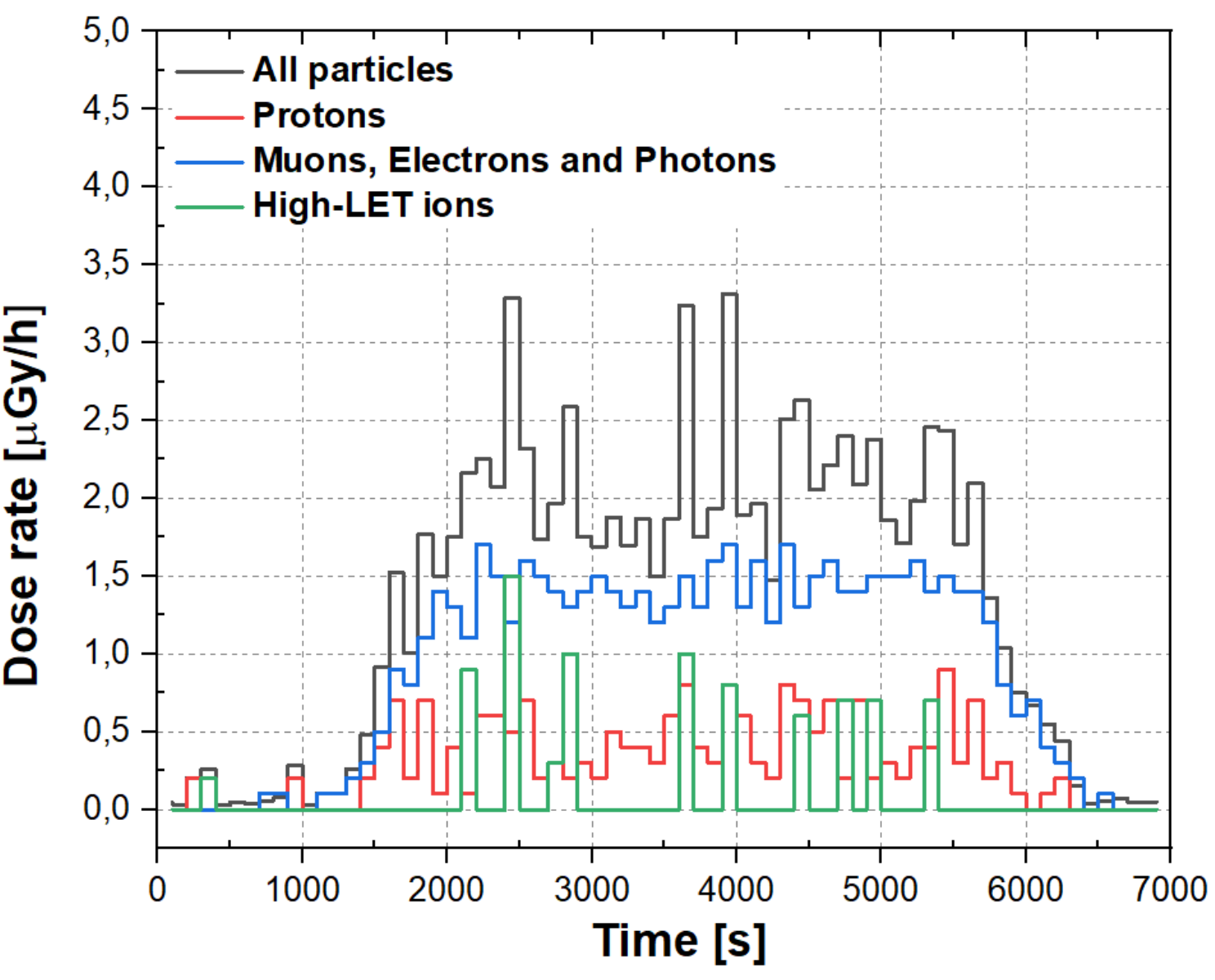}
 \caption{Particle flux (a) and dose rate (b) for identified particle classes including: protons, muons, photons and electrons, high LET-ions and other particles measured inside a commercial passenger plane at about 10 km altitude on 10$^{th}$ June 2019 on route Prague–Manchester. Data are given for the full sensor (total area 1.98 cm$^2$) scored for 100 s time bins along a continuous measurement of 115 min. Data collection started before take-off and ended after the plane landed.}
  \label{fig:flux_dose}
\end{center}
\end{figure}

ADVACAM s.r.o, Prague is a high-tech spin-off company from the Institute of Experimental and Applied Physics in Prague. ADVACAM commercializes Medipix/Timepix technology that is under development at CERN, Switzerland. The company has a long experience in silicon sensor manufacturing, micro packaging technologies, electronics design, software development and in applications of radiation imaging detectors. ADVACAM designs, manufactures and sells digital material-resolving color X-ray imaging cameras that represent the next generation of high-sensitivity and resolution imaging particle detectors. They are essentially cameras that register and display ionizing radiation of all types, such as X-rays, gamma-rays, electrons, ions, and even neutrons. Timepix-based detectors are characterized by high resolution, high scanning speed, and exceptional positional and directional sensitivity. These detectors can capture not only the particle's position and energy but also its path's shape. 
The Timepix technology and developed methodology have been used in various fields such as space radiation monitoring, radiation dosimetry, particle therapy for particle tracking, flux, dose and linear energy transfer spectra measurements, particle imaging, proton radiography detection and characterization of mixed radiation in high energy physics~\cite{GRANJA2014241,OANCEA2023102529, Oancea_2022,10.3389/fphy.2020.00346,Gohl_2022,Olsansky_2022,Granja_2022
}. With the use of a single chip detector, Timepix, on board an aircraft, it is possible to image and characterize the radiation field.
\begin{figure}[H]
\centering
 \includegraphics[width=0.5\textwidth]{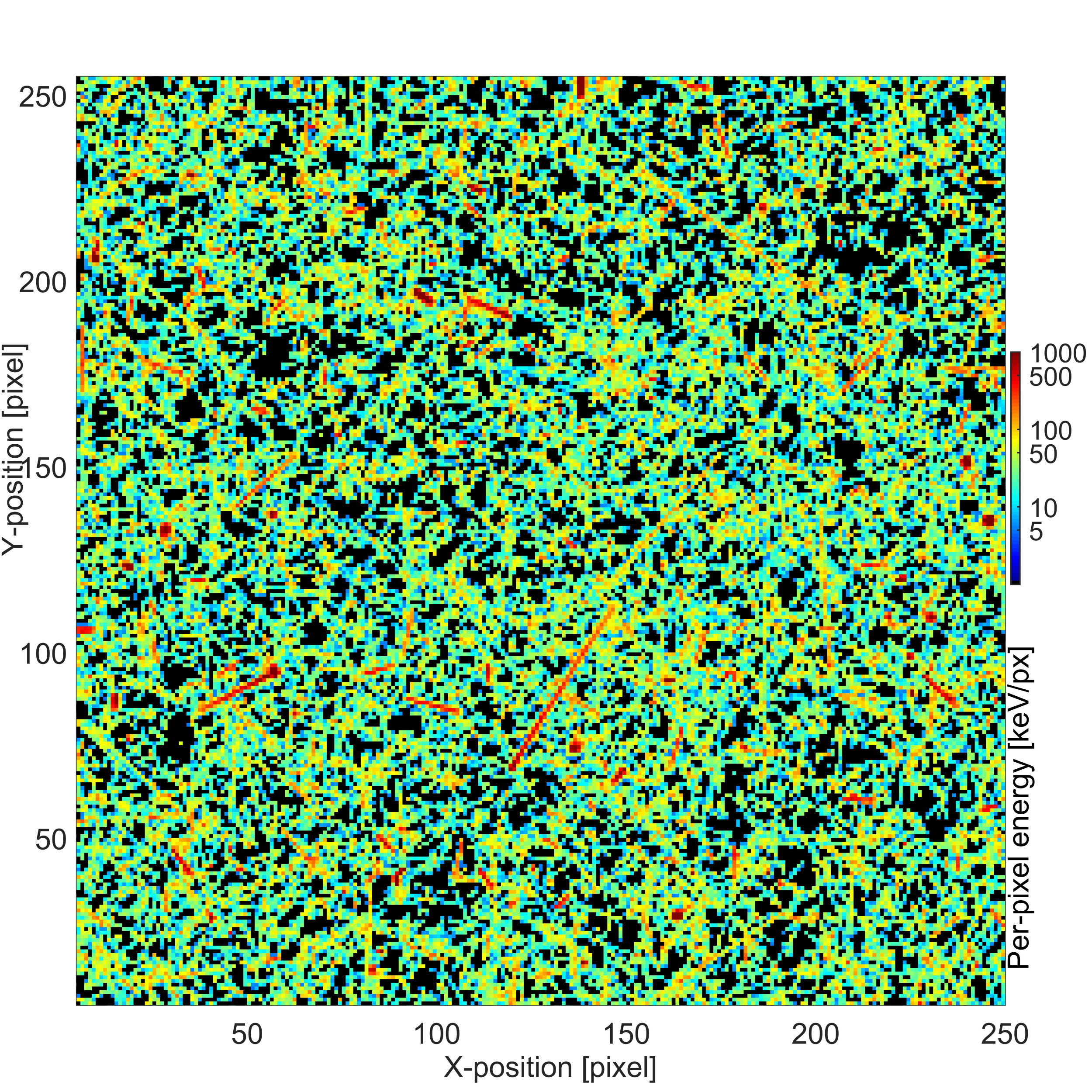}
 \caption{Integrated detection and 2D map of deposited energy in the Timepix detector (shown in color scale). Data collected during take off, at the flight altitude for the flight Prague-Manchester (6700s). The whole sensor pixel matrix is shown (256×256 pixels representing 14.08 ×14.08 mm$^2$).}
 \label{fig:all_data}
\end{figure}
Figure~\ref{fig:all_data} depicts the integrated detection and 2D map of deposited energy in the Timepix detector (shown in color logarithmic scale, where black means no particles were detected). The data were collected during the flight from Prague to Manchester (total of 6700s). The entire sensor pixel matrix is shown, which consists of 256×256 pixels representing 14.08 × 14.08 mm$^2$. As can be seen, there are registered tracks of various shapes, morphology and deposited energy. Based on cluster parameters, a classification of those tracks can be further made during post-processing. The identified classes of particles are protons, muons, photons, electrons, high linear energy transfer (LET) ions and other particles. The particle flux for individual classes can be seen in Figure~\ref{fig:flux_dose} (a), where the predominant class of detected particles is muons, photons and electrons. It is important to note that for this experiment calibrations were not performed, therefore in the reference fields the muons signals overlap with other recognized classes of particles: electrons and photons. On board the commercial aircraft the detected dose rate (DR) was 2.5673 uGy/h, whereas the measured DR at the ground level was 0.0927 uGy/h. The corresponding particle dose rates can be seen in Fig.~\ref{fig:flux_dose} (b). For a detailed description of the experimental setup, see~\cite{GRANJA2014241}.
\subsection{Astroteq}

\underline{Unique goal}: 

Astroteq is a startup that leverages AI and space technologies, it offers a unique solution to a problem that has plagued humankind for as long as it has existed - earthquakes. We are developing an earthquake forecast system that has the potential to impact by reducing the damages caused by earthquakes significantly, both financially and humanitarian. Our system utilizes deep machine learning techniques to analyze vast amounts of space weather and satellite data, identify patterns, find anomalies, and make forecasts that are impossible through traditional methods. The multichannel approach was undertaken by Astroteq and aims at implementing various measurements directly or indirectly connected to seismic activity.

\noindent \underline{Tech}:

The data of cosmic rays are time series of various modalities. Therefore, we built the multi-modal Convolutional Neural Network (CNN) architecture that predicts short-term and long-term data which models the dynamics features of multimodal on that time series data. Algorithms can automatically learn complex mapping of the data from input to output. It provides automatic learning of time dependence and trends appearing in data. We use LSTM and RNN structures that are effective in predicting the data inputs we have today (cosmic rays, weather data, weather forecast, seismic noise). However, they bear high computational cost.
\section{Outlook and conclusions}

In this document, we have presented several scientific problems that can be tackled within the CREDO Collaboration. The state-of-the-art studies have been briefly described whereas new, possible approaches have been overviewed. These include the inclusion of additional dedicated detectors to the smartphone CREDO net, the ongoing research on earthquake precursors, as well as the study of the Physics implications of detection of Cosmic Ray Ensembles.
\section{Acknowledgements}

D.E.A.C. acknowledges support from the NCN OPUS Project No. 2018/29/B/ST2/02576. P.H. is supported  by IRAP AstroCeNT (MAB/2018/7) funded by FNP from ERDF. The CREDO Project is generously supported by the Visegrad Fund under grant number 22220116.
\section{References}
\nocite{*}
\bibliographystyle{siam_bf}

\section*{Full Authors List: CREDO Collaboration}
%
%
\noindent
David Alvarez-Castillo$^{1,4}$,
Piotr Homola$^{1,26}$, 
Oleksandr Sushchov$^1$, 
Jaros\l{}aw Stasielak$^{1}$,
S\l{}awomir Stuglik$^{1}$,
Dariusz G{\'o}ra$^1$,
Vahab Nazari$^1$,
Cristina Oancea$^2$,
Carlos Granja$^2$,
Dmitriy Beznosko$^3$,
Noemi Zabari$^{4}$,
Alok C. Gupta$^5$,
Bohdan Hnatyk$^6$,
Alona Mozgova$^6$,
Marcin Kasztelan$^7$,
Marcin Bielewicz$^{7}$,
Peter Kovacs$^8$,
Bartosz {\L}ozowski$^9$,
Mikhail V. Medvedev$^{10,11}$,
Justyna Miszczyk$^1$,
{\L}ukasz Bibrzycki$^{14}$,
Micha\l{} Nied{\'z}wiecki$^{12}$,
Katarzyna Smelcerz$^{12}$,
Tomasz Hachaj$^{14}$
Marcin Piekarczyk$^{14}$,
Maciej Pawlik$^{13,14}$,
Krzysztof Rzecki$^{14}$,
Mat{\'i}as Rosas$^{15}$,
Karel Smolek$^{16}$,
Manana Svanidze$^{17}$,
Revaz Beradze$^{17}$,
Arman Tursunov$^{18}$,
Tadeusz Wibig$^{19}$,
Jilberto Zamora-Saa$^{20}$,
Bo\.zena  Poncyljusz$^{21}$,
Justyna M\k{e}drala$^{22}$,
Gabriela Opi{\l}a$^{22}$,
Jerzy Pryga$^{23}$,
Ophir Ruimi$^{24}$,
Mario Rodriguez Cahuantzi$^{25}$.
\\

\noindent
$^1$Institute of Nuclear Physics Polish Academy of Sciences, Radzikowskiego 152, 31-342 Krak{\'o}w, Poland.\\
$^2$ADVACAM, 12, 17000 Prague, Czech Republic\\
$^3$Clayton State University, Morrow, Georgia, USA.\\
$^4$Astrotectonic Ltd. (AstroTeq.ai), Juliusza Słowackiego 24, 35-069, Rzesz{\'o}w, Poland.\\
$^5$Aryabhatta Research Institue of Observational Sciences (ARIES), Manora Peak, Nainital 263001, India.\\
$^6$Astronomical Observatory of Taras Shevchenko National University of Kyiv, 04053 Kyiv, Ukraine.\\
$^7$National Centre for Nuclear Research, Andrzeja Soltana 7, 05-400 Otwock-{\'S}wierk, Poland.\\
$^8$Institute for Particle and Nuclear Physics, Wigner Research Centre for Physics, 1121 Budapest, Konkoly-Thege Mikl{\'o}s {\'u}t 29-33, Hungary.\\
$^9$Faculty of Natural Sciences, University of Silesia in Katowice, Bankowa 9, 40-007 Katowice, Poland.\\
$^{10}$Department of Physics and Astronomy, University of Kansas, Lawrence, KS 66045, USA.\\
$^{11}$Laboratory for Nuclear Science, Massachusetts Institute of Technology, Cambridge, MA 02139, USA.\\
$^{12}$Department of Computer Science, Cracow University of Technology, Warszawska 24, 31-155  Krak{\'o}w, Poland.\\
$^{13}$ACC Cyfronet AGH-UST, 30-950 Krak{\'o}w, Poland.\\
$^{14}$AGH University of Science and Technology, Mickiewicz Ave., 30-059 Krak{\'o}w, Poland.\\
$^{15}$Liceo 6 Francisco Bauz{\' a}, Lucas Obes 896, 11700 Montevideo, Uruguay.\\
$^{16}$Institute of Experimental and Applied Physics, Czech Technical University in Prague, Husova 240/5 110 00 Prague 1, Czech Republic.\\
$^{17}$E. Andronikashvili Institute of Physics under Tbilisi State University, 6, Tamarashvili St. 0186 Tbilisi, Georgia.\\
$^{18}$Institute of Physics, Silesian University in Opava, Bezru{\v c}ovo n{\'a}m. 13, CZ-74601 Opava, Czech Republic.\\
$^{19}$University of {\L}{\'o}d{\'z}, Faculty of Physics and Applied Informatics, 90-236 {\L}{\'o}d{\'z}, Pomorska 149/153, Poland.\\
$^{20}$Universidad Andres Bello, Departamento de Ciencias Fisicas, Facultad de Ciencias Exactas, Avenida Republica 498, Santiago, Chile.\\
$^{21}$Faculty of Physics, University of Warsaw, 02-093 Warsaw, Poland.\\
$^{22}$Faculty of Physics and Applied Computer Science, AGH University of Science and Technology, Mickiewicz Ave., 30-059 Krak{\'o}w, Poland Poland.\\
$^{23}$ Pedagogical University of Krakow, Institute of Computer Science, ul. Podchor\k{a}\.zych, 30-084 Krak{\'o}w, Poland.\\
$^{24}$Racah Institute of Physics, Hebrew University of Jerusalem, Jerusalem, IL, 91904, Israel\\
$^{25}$Facultad de Ciencias Físico Matemáticas-Benemérita Universidad Autónoma de Puebla, 72570, Mexico\\
$^{26}$AstroCeNT, Nicolaus Copernicus Astronomical Center Polish Academy of Sciences, ul. Rektorska 4, 00-614 Warsaw, Poland.
\end{multicols}
\end{document}